\documentclass[useAMS]{mn2e}

\usepackage{amssymb}
\usepackage{graphicx}
\usepackage{psfrag}
\usepackage{times,mathptm,mathrsfs,txfonts}
\makeatletter

\begin{document}

\input epsf
\title{Encounters between spherical galaxies I: \\ systems without dark halo}
\author[A.C. Gonz\'alez-Garc\'{\i}a \& T.S. van Albada]{A.C. Gonz\'alez-Garc\'{\i}a$^1$$^,$$^2$ and T.S. van Albada$^1$ \\
$^1$ Kapteyn Astronomical Institute,P.O. BOX 800, 9700 AV Groningen, The Netherlands\\
$^2$ Instituto de Astrof\'{\i}sica de Canarias, V\'{\i}a L\'actea s/n, La Laguna, 38200, Spain}

\maketitle
\begin{abstract}
We report here on a survey of N-body simulations of encounters between spherical galaxies. Initial systems are isotropic Jaffe models. Different sets of mass ratios, impact parameters and orbital energies are studied. Both merger remnants and systems perturbed after a non-merging encounter are analysed and compared to real-life elliptical galaxies. The properties of merger remnants show a large variety. Merger remnants resulting from head-on encounters are mainly non-rotating prolate spheroids. Merger remnants from models with ${\bf J}_{\rm orb} \neq 0$ are triaxial or mildly oblate spheroids supported in part by rotation. The velocity distributions are biased toward the radial direction in the prolate case and the tangential direction in the oblate case. Non-mergers are affected in various ways, depending on the orbital characteristics. We conclude that many of the global properties of real-life ellipticals can in principle be attributed to a merger of spherical progenitors.
\end{abstract}

\section{Introduction}

Elliptical galaxies, for a long time believed to be simple rotating spheroidal systems, are now known to be far more complex. From the study by Bertola \& Capaccioli (1975) we know that support against gravity is provided by random motions as well as rotation. Remarkably the surface brightness distribution of ellipticals closely follows the $r^{1/4}$ law proposed by de Vaucouleurs (1958), although a more general $r^{1/n}$ profile, with $1<n<10$, (S\'ersic 1968) appears to provide a somewhat better fit (Caon et al.~ 1993, Bertin et al.~2002). Several dynamical models have been developed trying to reproduce this $r^{1/4}$ shape (e.g. van Albada 1982). The shape of elliptical galaxies is in general triaxial and goes from oblate to prolate (de Zeeuw \& Franx 1991). With few exceptions elliptical galaxies contain only small amounts of gas, and can be considered as dissipationless systems. 

At present there exist two main views on how elliptical galaxies are formed. On the one hand a huge gas cloud may contract rapidly and form stars, resulting in an elliptical like galaxy. This is known as the gravitational collapse theory. On the other hand elliptical galaxies might form via assembly of small building blocks: encounters of small systems and the subsequent mergers will give rise to larger galaxies. Elliptical galaxies formed through this process could even be the result of mergers of disc galaxies. This is known as the hierarchical merging theory. 

Since the early work by Toomre \& Toomre (1972) pointing towards merging of spiral galaxies as a likely key factor in galaxy evolution, a large effort has been devoted not only to observe interactions and mergers of galaxies but also to model these  processes and the resulting systems. On the observational side, features found in such detailed studies include: tidal tails, bridges, shells and counter-rotation (see Schweitzer 1986 and Schweitzer 1998 for reviews of the observations). Modelling of such features in individual systems with N-body simulations has in general been quite successful (see Barnes 1999 and Naab \& Burkert 2001 for recent reviews, and for the various features: Hernquist and Quinn 1988, Merrifield and Kuijken 1998, Balcells and Quinn 1990, Hernquist and Barnes 1991, Balcells and Gonz\'alez 1998, Toomre 1978, Higdon 1995, Weinberg 1998, Garc\'{\i}a-Ruiz et al.~2001).

Early papers on interactions of spheroidal systems, (van Albada \& van Gorkom 1977, White 1978, White 1979, Farouki et al.~1983, Aguilar \& White 1985) focus on the exchange of mass and energy. The main aim of these and of several later studies was to explore the conditions under which a merger will take place and to probe the general morphology of the merger remnants. For example, Navarro, in a series of papers (Navarro ($1989$), Navarro \& Mosconi ($1989$) and Navarro ($1990$)), find that merger remnants resulting from an encounter of non-rotating de Vaucouleurs models have a tail in the light distribution departing from  the $r^{1/4}$ law. These simulations, with a few exceptions, have been done with one thousand particles at most. They therefore lack the resolution needed for a detailed study of the kinematics of the endproducts. In recent years mergers of spheroidal galaxies have not received much attention, although a number of studies continue along the lines of earlier work (Okamura et al.~1991, Vergne \& Muzzio 1995, Combes et al.~1995, Levine \& Aguilar 1996, Seguin \& Dupraz 1996 and Makino and Hut 1997). Detailed study of the structure of the merger remnants, in particular the kinematics, is still largely unexplored. In this respect the structure of merger remnants resulting from encounters between spheroidal galaxies is still poorly known. 

Until recently the merger rate of galaxies was crudely known. Therefore it has been difficult to put the results of the work described above into the context of galaxy evolution. As shown recently by van Dokkum et al.~(1999) in their study of clusters at medium and high redshift, a large fraction of elliptical (or spheroidal) systems are undergoing mergers (up to $50 $ per cent). Therefore, the study of the end products of encounters between spheroidal galaxies is important for our understanding of elliptical galaxies. This has led us to the following questions: (1) what is the relation between the input parameters describing an encounter (masses, impact parameter, relative velocity) and the parameters describing the merger remnant (shape, kinematics)? (2) how do the parameters of the merger remnants compare with those of real-life ellipticals?

To answer these questions we have carried out simulations of collisions between spherical systems with a variety of input parameters. We chose three input parameters to keep our experiments  simple but at the same time reasonably complete. The models were let to evolve with different orbital conditions and with different mass ratios between the progenitor systems. The results are presented in the form of a review of the main characteristics of the end products. Results from these simulations concerning the evolution of the Fundamental Plane of Elliptical galaxies due to merging can be found in Gonz\'alez-Garc\'{\i}a \& van Albada (2003). In Paper II we include the effects of the presence of a dark matter halo and compare with the results obtained here.

The end products of these merger simulations show a large variety of properties. Their characteristics can be traced to specific initial conditions. For some features the impact parameter is the key to the final configuration, for other features it is the initial mass ratio of the progenitor systems. Either way we find that one may attribute several characteristics of elliptical galaxies to collisions between `simple', i.e. spherical progenitor galaxies.

We conclude that the various characteristics of elliptical galaxies as observed now are not detroyed by mergers. In fact mergers between spheroidal systems (as well as discs) may well be resposible for many of these characteristics.

\section{Models}

Below we describe the setup of our experiments and the integration method.

\subsection{Initial conditions}

	As initial conditions we use purely stellar non-rotating isotropic spherical systems described by Jaffe's law (Jaffe 1983). The implementation used here was developed by Smulders \& Balcells (unpublished). The distribution function (DF) is derived from the potential and the density, and the particle positions and velocities are obtained from the DF.

For a Jaffe model the potential is given by:

\begin{equation}
	\phi(r) = \frac{GM}{r_{\rm J}}\,{\rm \ln} \left(\frac{r}{r+r_{\rm J}}\right) ,
\end{equation}
where G is Newton's constant of gravity, {\it M} is the total mass of the system and $r_{\rm J}$ is the half mass radius. The corresponding mass density is:

\begin{equation}
	\rho(r) = \left(\frac{M}{4\pi r_{\rm J}^3}\right)\frac{r_{\rm J}^4}{r^2(r+r_{\rm J})^2} 
\end{equation}
and the mass inside radius {\it r} is:

\begin{equation}
M(r) = \frac{r}{r+r_{\rm J}} M .
\end{equation}
Further,

\begin{equation}
\langle v^2 \rangle = \frac{GM}{2r_{\rm J}} .
\end{equation}

For a spherical system with an isotropic velocity distribution the DF is a function of energy only (Binney \& Tremaine $1987$). For Jaffe's model the potential-density pair has an analytical solution for the DF given by:

\begin{eqnarray}
\label{DF}
	\lefteqn{f(E) =\frac{M}{2\pi^3(GMr_{\rm J})^{3/2}}} \\ 
& & \left[F_- \left(\sqrt{2 \mathscr{E}} \right)-\sqrt{2}F_-\left(\sqrt{ \mathscr{E}}\right) - \sqrt{2} F_+\left(\sqrt{ \mathscr{E}}\right)+F_+\left(\sqrt{2 \mathscr{E}}\right)\right], \nonumber
\end{eqnarray}
where $\mathscr{E} \equiv (-Er_{\rm J}/GM)$ and $F_{\pm}(x)$ is Dawson's integral: $F_{\pm}(x)\equiv e^{\mp x^2}\int^x_0 dx e^{\pm x^2}$ (Jaffe 1983).
Particles are placed in a sphere with positions and velocities provided by the above DF. 

When constructing the models a cut-off radius is imposed to avoid having particles at exceedingly large radius.  The cut-off radius is equal to {$10 \times r_{\rm J}$}, corresponding to the $91\%$ mass radius of the Jaffe model. This results in a small change of the structural parameters. By cutting off the outter $9 \%$ of the mass distribution the half-mass radius of the model is reduced to the $82 \%$ of the theoretical $r_{\rm J}$. 

\subsection{Units \label{units}}

Non-dimensional units are used throughout. We adopt $G=1$ for Newton's constant of gravity. The theoretical half mass radius of the Jaffe model  $r_{\rm J}$  and the total mass $M$  of the smaller system are also set equal to 1. Then, for a Jaffe model $\langle v^2 \rangle = 1/2$ and the half-mass crossing time $T_{\rm cr} = 2 r_{\rm J} /\langle v^2 \rangle^{1/2} = 2\sqrt{2}$. The models may be compared with real galaxies using, for example, the following scaling:

\begin{equation}
	        [M] = M_{\rm J} = 4\times10^{11} \; \rmn{M_{\odot}}, \\
\end{equation}
\begin{equation}
		[L] = r_{\rm J} = 10 \;\rmn{kpc}  ,\\
\end{equation}
\begin{equation}
		[T] =  2.4\times10^7 \;\rmn{yr}. \\
\end{equation}
With these, the velocity unit is:

\begin{equation}
		[v] =  414\; \rmn{km/s}. \\
\end{equation}

\subsection{Method \label{method}}

We have used Hernquist's ($1987$, $1990$) version of the {\small TREECODE} on an Ultra-Sparc station where a typical run (with $2 \times 10^4$ particles) takes of the order of $10^5$ seconds for about $10^4$ timesteps.  Softening was set to $1/8$ of the half mass radius of the smallest galaxy ($R_{1/2}=0.82$ and $\varepsilonup = 0.1$). The tolerance parameter was set to $\theta=0.8$. We have done a test run with $\theta=0.6$. No major differences were noticed, while the computational time is increased with a factor 2. Quadrupole terms were included in the force calculation and the time step was set to $1/100$ of the half mass crossing time.

\subsection{Stability of initial models}

\begin{figure}

\begin{center}


\vbox{%

\epsfxsize=8cm

\epsfbox{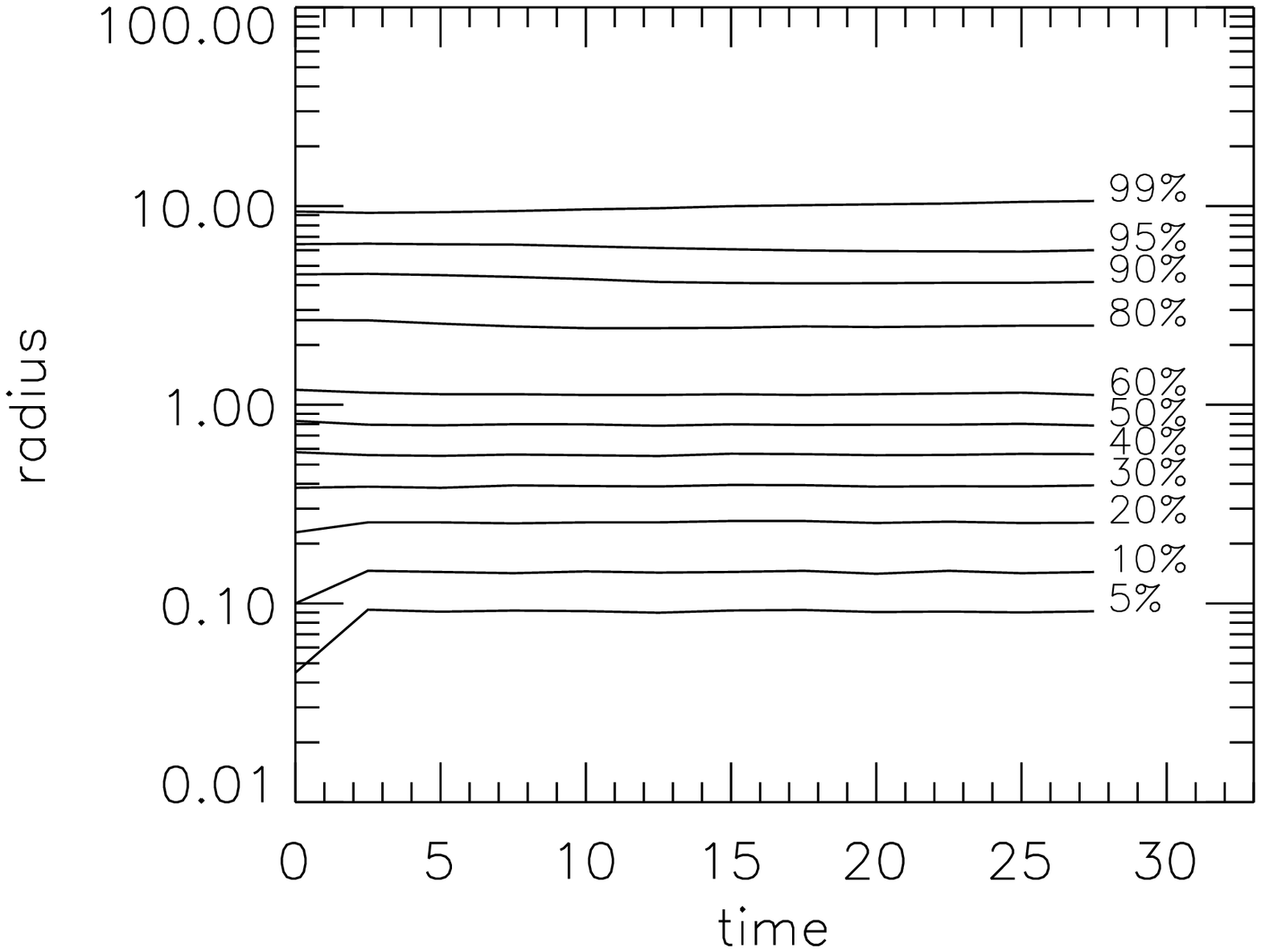}

\epsfxsize=8cm

\epsfbox{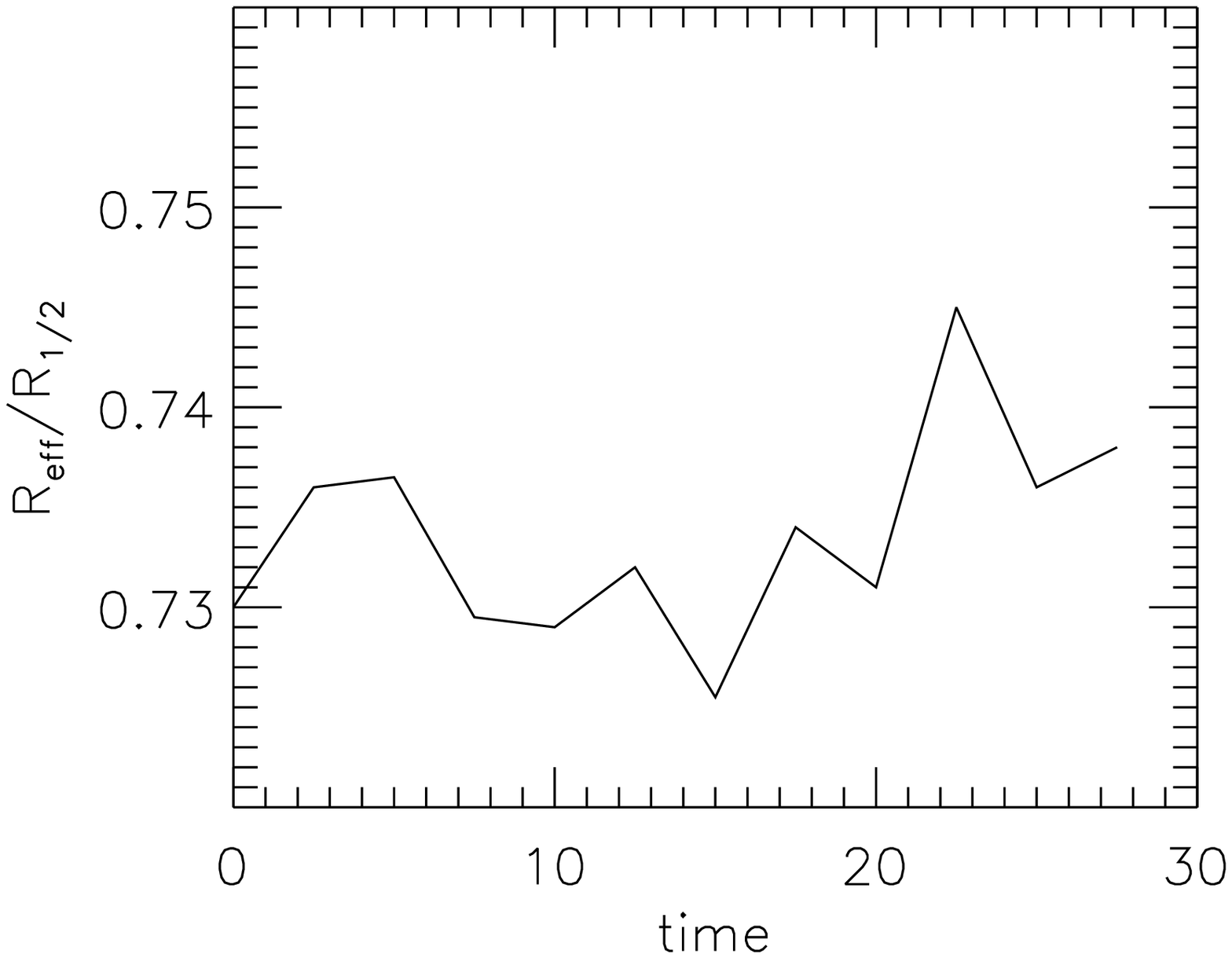}}

\caption{A test run was made to test the stability of our initial models. In the top panel we give  the evolution of the mass inside different radii, the top line gives the $99 $ per cent mass radius, while the bottom one gives the $5  $ per cent mass radius. The lower panel gives the ratio between the half mass radius ($R_{1/2}$) and the effective radius ($R_{\rm e}$). For a Jaffe model this ratio should be 0.74.}

\label{fig:test1}

\end{center}


\end{figure}

We have checked the stability of our input initial model for $28$ time units, which correspond to 10 crossing times (see section \ref{units}). We use a model with 10240 particles and a softening, $\varepsilonup = 0.1$. The test shows (see Figure \ref{fig:test1} top panel) that the system relaxes for about 4 time units and remains stable thereafter. This initial relaxation is due to the presence of the particle softening in the code. 

The ratio of the effective radius ($R_{\rm e}$, measured as the radius enclosing half of the mass in projection) and the half mass radius ($R_{1/2}$) remains close to the value $0.74$ given by Dehnen ($1993$) for a pure Jaffe model; see Figure \ref{fig:test1} bottom.

\subsection{Initial parameters}

We vary the following three input parameters: the initial orbital energy, the impact parameter of the orbit and the mass ratio of the two galaxies.

Let {\it M}, {\it R} and $\sigma = \langle v^2 \rangle ^{1/2}$ denote mass, radius and velocity dispersion. According to the virial theorem the velocity dispersion is given by $\sigma^2 = \alpha GM/R$ where $\alpha$ is a structure constant appropriate for a Jaffe-law system. Models with different masses were constructed following the scaling relation between mass and radius given by Fish (1964), i.e.:

\begin{equation}
M_1/R^2_1 = M_2/R^2_2 = constant.
\label{VT}
\end{equation}
Homologous systems then also follow the scaling relation:

\begin{equation}
\sigma_1^2/R_1 = \sigma^2_2/R_2.
\label{eqvt2}
\end{equation}

To calculate the initial orbital energy we proceed as follows: the total energy of the system is given by the sum of the internal energy of each system and the relative orbital energy:

\begin{eqnarray}
\lefteqn{E_{\rm tot}= E_{\rm i1} + E_{\rm i2} + E_{\rm orb} }\\
& & =-\frac{1}{2} M_1\sigma^{2}_1 -\frac{1}{2}M_2\sigma^{2}_2-\frac{GM_1M_2}{R}+\frac{1}{2} \frac{M_1M_2}{M_1+M_2} V^2,\nonumber
\end{eqnarray}
where $E_{\rm i1}$ and $E_{\rm i2}$ are the internal energies of galaxies $1$ and $2$ and $E_{\rm orb}$ is the orbital energy. Consider the case where at infinity the total energy of the system is zero. Then,

\begin{equation}
\label{eq:intener}
-\frac{1}{2} M_1\sigma^{2}_1 -\frac{1}{2}M_2\sigma^{2}_2+\frac{1}{2} \frac{M_1M_2}{M_1+M_2} V_{\infty}^2(0) = 0,
\end{equation}
where $V_{\infty}(0)$ denotes the relative velocity at infinity for a system with zero total energy.

\begin{table*}
\begin{center}
\begin{minipage}{120mm}
\caption{Input parameters. The columns identify the run and give the mass ratio, impact parameter and orbital energy.\label{tab1}}
\begin{tabular}{@{}cccccccc}
\hline
{\bf Run} &{\bf $M_2:M_1$}& {\bf $Impact Par.$} & {\bf $E_{\rm orb}$} & {\bf Run} &{\bf $M_2:M_1$}& {\bf $Impact Par.$} & {\bf $E_{\rm orb}$} \\
\hline
$1hP$ & 1:1 & 0 & 0 & $5hP$ & 5:1 & 0 & 0\\
$1hH$ & 1:1 & 0 & 0.0625 & $5hH$ & 5:1 & 0 & 0.761\\
$1hZ$ & 1:1 & 0 & 0.250 & $5hZ$ & 5:1 & 0 & 1.522\\
$1hW$ & 1:1 & 0 & 0.360 & & & & \\
$1oP$ & 1:1 & 5 & 0 & $5oP$ & 5:1 & 11.18 & 0\\
$1oH$ & 1:1 & 5 & 0.0625 & $5oH$ & 5:1 & 11.18 & 0.761 \\
$1oZ$ & 1:1 & 5 & 0.250 & $5oZ$ & 5:1 & 11.18 & 1.522\\
$1gP$ & 1:1 & 10 & 0 & $5gP$ & 5:1 & 22.36 & 0 \\
$1gPG$ & 1:1 & 10 & 0 &  &  &  &  \\
$1gH$ & 1:1 & 10 & 0.0625& $5wW$ & 5:1 & 1.14 & 1.448\\
$1wP$ & 1:1 & 15 & 0 & & & & \\
$2hP$ & 2:1 & 0 & 0 & $7hP$ & 7:1 & 0 & 0\\
$2hH$ & 2:1 & 0 & 0.239 & $7hH$ & 7:1 & 0 & 1.22\\
$2hZ$ & 2:1 & 0 & 0.478 & $7hZ$ & 7:1 & 0 & 2.44\\

$2oP$ & 2:1 & 7.07 & 0 & $7oP$ & 7:1 & 13.23 & 0\\
$2oPr$ & 2:1 & 5 & 0 & & & & \\
$2oHr$ & 2:1 & 5 & 0.239 & $7oH$ & 7:1 & 13.23 & 1.22\\
$2oZr$ & 2:1& 5 & 0.478&  &  &  & \\
$2gP$ & 2:1 & 14.14 & 0 & $7gP$ & 7:1 & 26.46 & 0\\
$2gPr$ & 2:1 & 10 & 0 &  &  &  & \\
$2gHr$ & 2:1 & 10 & 0.239 & & & &\\
$3hP$ & 3:1 & 0 & 0 & $10hP$ & 10:1 & 0 & 0\\
$3hH$ & 3:1 & 0 & 0.387 & $10hH$ & 10:1 & 0 & 2.039\\
$3hZ$ & 3:1 & 0 & 0.775 & $10hZ$ & 10:1 & 0 & 4.078\\
$3oP$ & 3:1 & 8.66 & 0 & $10oP$ & 10:1 & 15.81& 0\\
$3oH$ & 3:1 & 8.66 & 0.387 & $10oH$ & 10:1 & 15.81 & 2.039 \\
$3oZ$ & 3:1 & 8.66 & 0.775 &&&&\\
$3gP$ & 3:1 & 17.32 & 0 & $10gP$ & 10:1 & 31.62 & 0 \\
$3gH$ & 3:1 & 17.32 & 0.387 &&&&\\
\hline  
\end{tabular}
\end{minipage}
\end{center}
\end{table*}

Given $M_1$, $M_2$, $\sigma_1$ and $\sigma_2$ we can calculate the corresponding value of $V_{\infty}(0)$. Using equations \ref{VT}, \ref{eqvt2} and \ref{eq:intener} the velocity at infinity can be expressed as a function of the velocity dispersion of the smallest system at the beginning, $\sigma_1$:

\begin{equation}
V_{\infty}^2(0)=\left(1+\frac{M_1}{M_2}\right)\left(1+\left(\frac{M_1}{M_2}\right)^{-3/2}\right) \sigma_1^2
\end{equation}

In order to represent a range of energies, we have chosen three different values for $V_{\infty}$, given by: $V_{\infty}^2=0$, $\;0.5 V_{\infty}^2(0)$, and $\;V_{\infty}^2(0)$ respectively.

With this choice for the orbital energy, $E_{\rm orb}\geq 0$, we have parabolic  ($V_{\infty} =0$) as well as hyperbolic encounters. Part of the orbital energy, both in parabolic and hyperbolic encounters, may be lost in the subsequent interactions of the two systems, leading to a bound system with $E_{\rm orb}< 0$, in those cases where the interaction is strong enough. So  cases with $E_{\rm orb}< 0$, although not included explicitly as starting points,  will appear as a consequence of interaction.

The impact parameter ($D$) of the orbit was chosen as follows: we are interested in head-on collisions, $D=0$, as well as encounters with $D\neq0$, but with $D<R_{\rm out}$, where $R_{\rm out}$ is the radius including $99\%$ of the mass of the most massive galaxy. Impact parameters $D>R_{\rm out}$ would lead to a weak interaction which is hard to follow numerically. We selected three values of $D$:  $D=0$ (head-on collision), $D=R_{\rm out}/2$ and $D=R_{\rm out}$. To define better the border line between merger and non-mergers, some models with mass ratio 2:1 and $D \neq 0$ have also been run with initial values of $D$ slighlty smaller than the scheme presented above. 

The parameters of the initial models are summarized in Table \ref{tab1},  where the units used are model units. The first column gives the model name. The number indicates the mass ratio, the small letter denotes the impact parameter, with {\it h} for head-on ($D=0$), {\it o} for off set ($D=R_{\rm out}/2$) and {\it g} for grazing encounters ($D=R_{\rm out}$). The final capital letter denotes the energy of the orbit with {\it P} for parabolic, {\it H} for hyperbolic and {\it Z} for zero energy at infinity. The models with mass ratio 2:1 and slightly different $D$ are labelled with an $r$.
For some specific cases we have chosen slightly different initial conditions, these are given by a letter {\it w} in the name (runs $1hW$, $1wP$ and $5wW$) .

For most runs we use $10240$ particles per galaxy. To avoid heating due to large differences in particle masses we used 25600 particles for the most massive galaxies ($M_2 \geq 7$).  Model $1gP$ was repeated with 10 times more particles giving basically the same results as can be seen in the following sections. This is model $1gPG$.

Models leading to a merger were evolved for at least $8$ to $10$ dynamical crossing times of the merged system after merging, to allow the system to relax (reach virialization). Conservation of energy is good in all the runs; variations are less than $0.5$ per cent.

\section{Results}

\begin{figure}

\begin{center}
\leavevmode
\hbox{%

\epsfxsize=9cm
\epsfbox{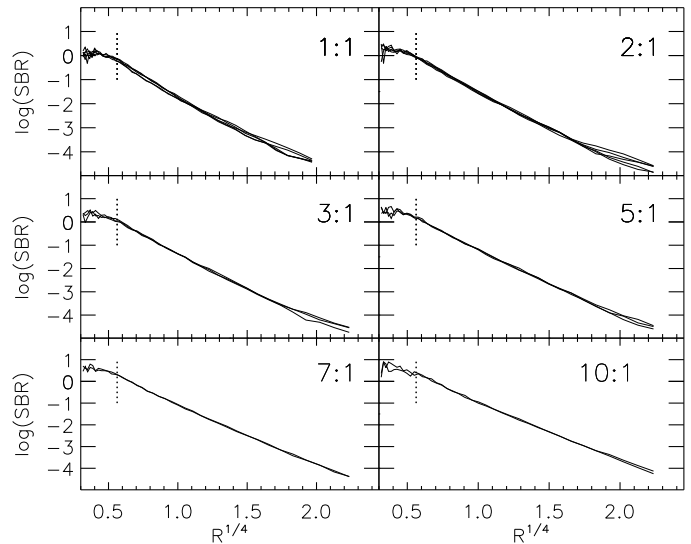}}
\caption{The surface brightness of the merger remnants versus $R^{1/4}$. In this plot a de Vaucouleurs (1958) profile is given by a straight line. Note the slight upturn with respect to a de Vaucouleurs' profile in the outer regions.  The dotted line indicates the softening lenght. \label{devau}}
\end{center}
\end{figure}

The present sample consist of 49 runs; the results  are summarized in Table \ref{tab:Enohres} and are discussed below. The merger remnants do resemble real-life elliptical galaxies. Figure \ref{devau} shows their surface density profiles. These can be fitted by a de Vaucouleurs (1958) profile for about 10 magnitudes. The innermost points fall below a de Vaucouleurs profile, this result is suspect however because these points lie inside one softening radius ($\varepsilonup^{1/4} \sim 0.56$).

In the presentation below we focus on properties that can be compared with observations.

\begin{figure*}

\begin{center}
\leavevmode
\hbox{%

\epsfxsize=14cm
\epsfbox{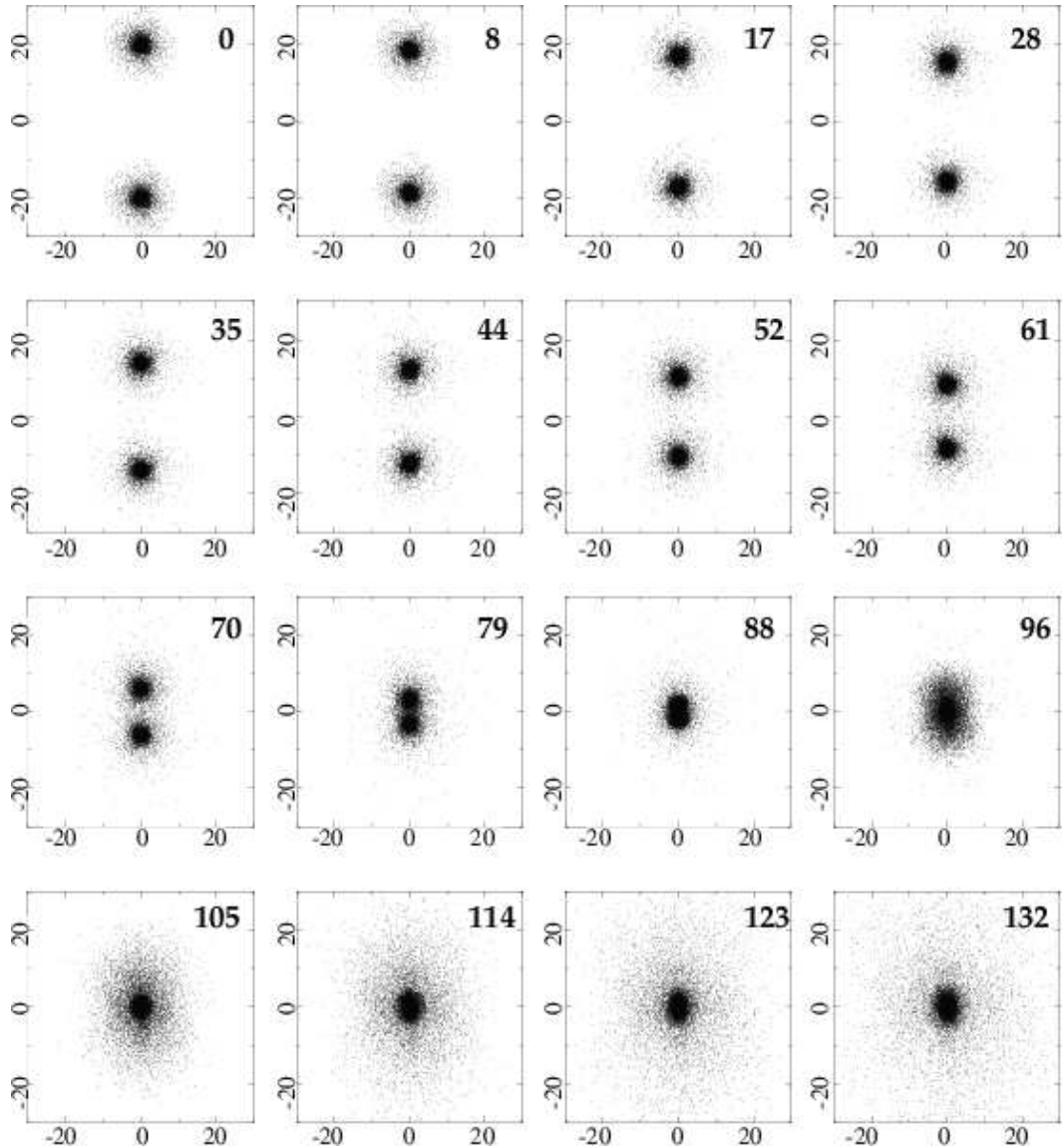}}
\caption{Evolution of systems in run $1hP$. This is a head-on collision between two equal mass galaxies. Numbers at the top of each frame show the time in computational units. The first encounter occurs around time 85. The two systems develop a faint `halo' after this first encounter. Many particles in these haloes will eventually escape. The end result is a prolate cigar-shaped E5 system.\label{fig:j11}}
\end{center}
\end{figure*}

\begin{figure*}

\begin{center}
\leavevmode
\hbox{%

\epsfxsize=14cm
\epsfbox{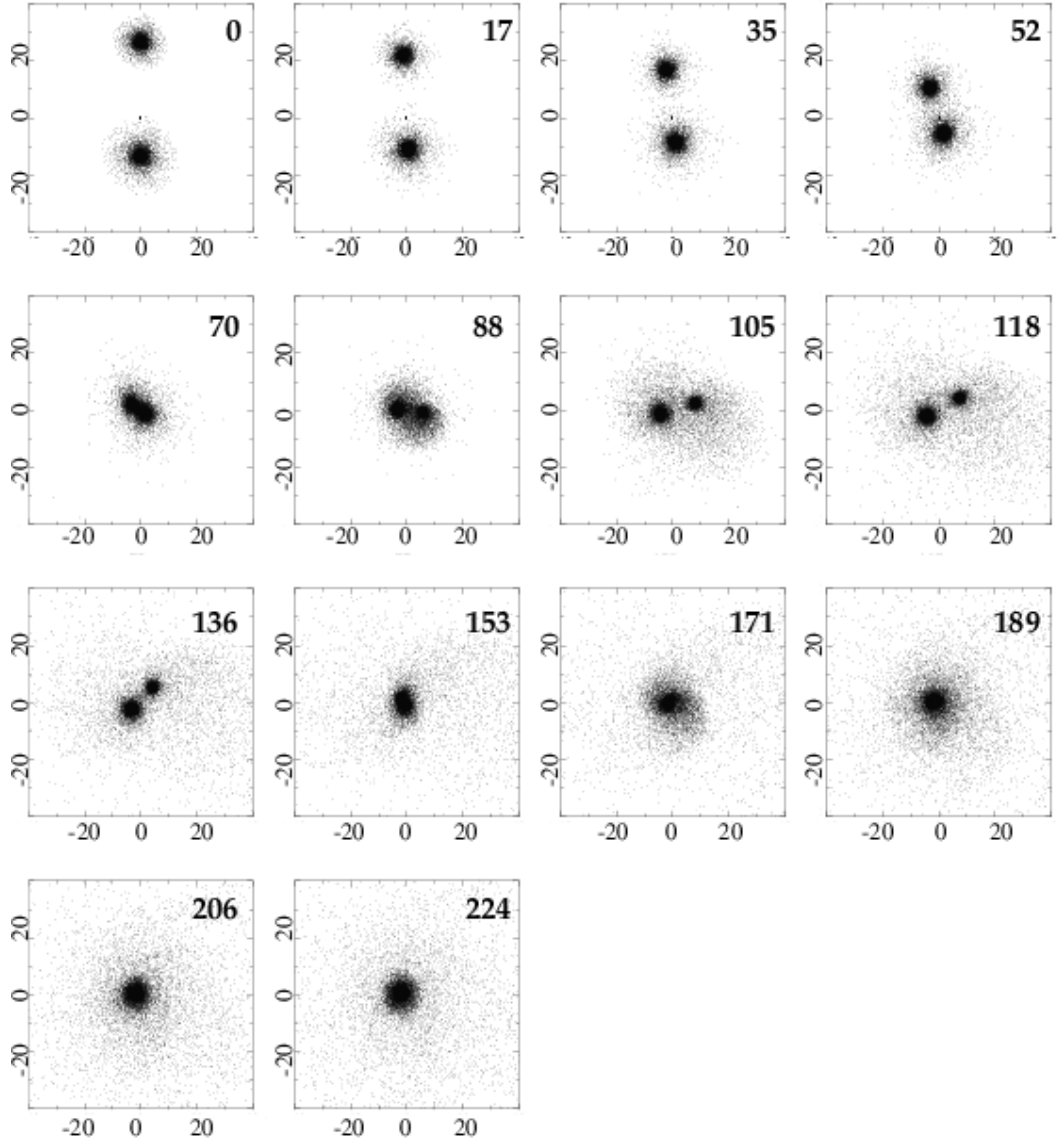}}
\caption{Evolution of the systems in run $2gPr$. This is an encounter between two systems with mass ratio 2:1, where the galaxies meet in a parabolic orbit with impact parameter slightly smaller than the outer radius of the largest galaxy (initially at bottom). The final system develops a rotating bar.\label{fig:j12D}}
\label{fig:j12D}
\end{center}
\end{figure*}

\begin{figure*}

\begin{center}
\leavevmode
\hbox{%

\epsfxsize=13cm
\epsfbox{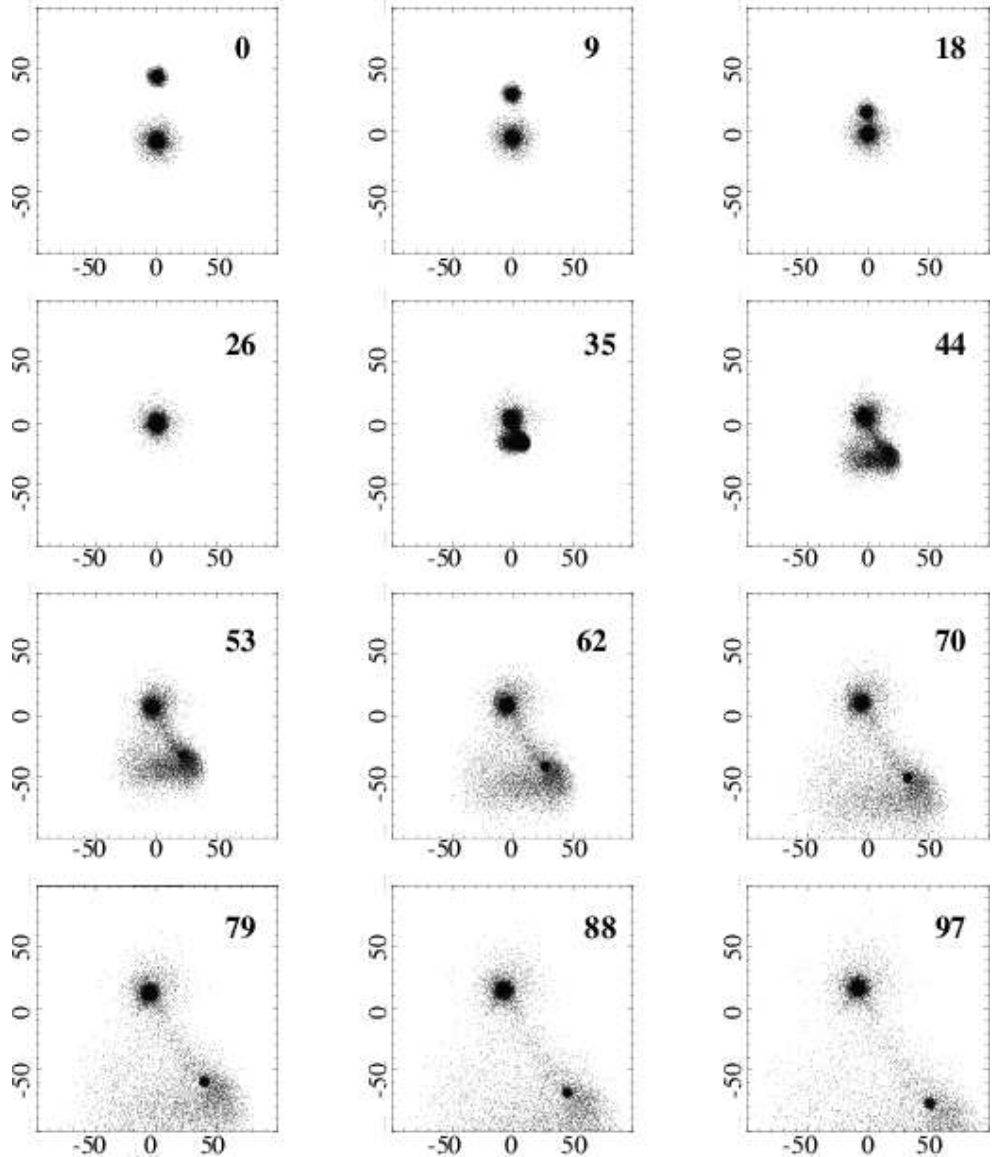}}
\caption{Evolution in run $5w$. The initial systems are placed on a grazing hyperbolic orbit. This simulation does not end in a merger. Note the plume of particles ahead and behind the smaller system.\label{fig:wrong}}
\label{fig:wr}
\end{center}
\end{figure*}

\subsection{Phenomenology}

\begin{table*}
\begin{center}
\begin{minipage}{123mm}
\caption{Final systems. Column (1) identifies the run and (2) gives the time where the run was stopped. (3) indicates whether the final result is a merger (Y) or not (N). Models which might lead to a merger but will take too long to follow computationally are denoted with `??'. (4) provides the l.o.s. ellipticity for the isophote at $R_{\rm e}$ ($\epsilon$), and (5) and (6) give the axis ratios $b/a$, and $c/a$. In (7) the ratio between the  rotational velocity and the central velocity dispersion is given. For those models not leading to merger two measurements are included, the first one is for the less massive system and the second for the more massive one. 
\label{tab:Enohres}}
\begin{tabular}{ccccccc}
\hline
{\bf Run} &{\bf $t_{fin}$}&  {\bf Merger} & {\bf $\epsilon$}& { \bf $b/a$}& {\bf $ c/a$} & {\bf $V_{\rm max}/\sigma_0$} \\
(1) & (2) & (3) & (4) & (5) &  (6) & (7) \\
\hline
$1hP$ & 110 &	 $Y$ & 0.528 &0.771 &0.750 & 0.09 \\
$1hH$ & 123.2 & $Y$ & 0.445 &0.812 &0.750 & 0.16 \\
$1hZ$ & 167.2 & $Y$ &  0.449 & 0.788& 0.746& 0.107 \\
$1hW$ &110 & $N$ &0.196//0.169 &0.966//0.935 &   0.946//0.896 &0.304//0.292  \\
$1oP$ & 154 & $Y$ & 0.394 & 0.928& 0.816&  0.615 \\
$1oH$ & 149.6 & $N$ & 0.164//0.206 &0.990//0.970 & 0.9656//0.958& 0.42//0.29 \\
$1oZ$ &110 & $N$ &  0.125//0.141&0.995//0.976 &0.992//0.972 & 0.19//0.19 \\
$1gP$ & 219.9 & $Y$ & 0.268 &0.953 &0.749 & 0.638 \\
$1gPG$& 245 & $Y$ & 0.312 &0.967 &0.736 & 0.592 \\
$1gH$ & 110 & $N$ & 0.236//0.220&0.978//0.992 & 0.960//0.968& 0.12 // 0.2 \\
$1wP$ & 220& $??$ &0.583//0.452  &0.979//0.982 & 0.966//0.964&0.146//0.120  \\
$2hP$ &  110 & $Y$ & 0.332 &0.813 &0.795 & 0.18 \\
$2hH$ &  136.4 &  $Y$ & 0.264 &0.840 &0.783 & 0.15  \\
$2hZ$ & 219.9 &  $Y$ & 0.231 &0.844 &0.816 & 0.15 \\
$2oP$ & 172 & $Y$ & 0.314&0.977 &0.844 & 0.45 \\
$2oPr$ & 110 & $Y$ & 0.247 &0.926 &0.828 & 0.45 \\
$2oHr$ &  110 &  $N$ & 0.094//0.149&0.983//0.990 &0.958//0.975 & 0.18//0.24 \\
$2oZr$ & 110 &  $N$ &  0.104//0.106&0.972//0.977 &0.963//0.964 & 0.26//0.17 \\
$2gP$ & 600 & $Y-??$ & 0.312 &0.963 &0.765 & 0.61 \\
$2gPr$ & 224.3 & $Y$ & 0.259 &0.984 &0.803 & 0.77 \\
$2gHr$ & 162.8 & $N$ & 0.16//0.18 &0.990//0.987 &0.964//0.951 & 0.19//0.34 \\
$3hP$ & 110 & $Y$ & 0.250 &0.841 &0.827 & 0.246 \\
$3hH$ & 110 & $Y$ & 0.203 &0.901 &0.880 & 0.195 \\
$3hZ$ & 110 & $N$ & 0.184//0.081&0.987//0.990 &0.959//0.963 & 0.2//0.25 \\
$3oP$ & 176 & $Y$ & 0.196 &0.954 &0.853 & 0.436 \\
$3oH$ & 110 & $N$ & 0.103//0.101&0.9689//0.972 &0.937//0.956 & 0.18//0.27 \\
$3oZ$ & 110 & $N$ & 0.104//0.102 &0.990//0.987 &0.968//0.957 & 0.19//0.38 \\
$3gP$ & 110 & ?? &0.300//0.237 &  0.988//0.972 &  0.969//0.958 &0.232//0.292 \\
$3gH$& 110 & $N$ & 0.158//0.125&0.979//0.968 &0.939//0.965 & 0.17//0.35 \\
$5hP$ & 123.2 & $Y$ & 0.152 &0.875 &0.853 & 0.34\\
$5hH$ & 110 &$Y$ &0.107 &0.778 &0.777 & 0.296 \\
$5hZ$ & 110 & $N$ & ---//0.078 &0.923//907 &0.827//0.905 & ---//0.222\\
$5oP$ & 219.9 &$Y$ & 0.127 &0.983 &0.909 &  0.42 \\
$5oH$ & 110 & $N$ &0.176//0.114 &0.984//0.994 &0.963//0.977 & 0.14//0.36 \\
$5oZ$&158.4 &$N$ &0.135//0.114&0.977//0.989 &0.964//0.951 & 0.23//0.23 \\
$5gP$ &206.7 &?? &0.290//0.197 &0.973//0.985 &0.956//0.981 & //0.203\\  
$5wW$ &101 &$N$ &0.078//0.139&0.975//0.967 &0.967//0.939 & 0.36//0.28 \\
$7hP$ & 110 & $Y$& 0.103&0.953 &0.943 & 0.21\\
$7hH$ & 110 & $N$&---//0.106&0.869//0.991 &0.862//0.976 & ---//236\\ 
$7hZ$ & 110 & $N$&---//0.060&---//0.996 & ---//0.983 & ---//0.200 \\
$7oP$ & 334 & $Y$& 0.109&0.993 &0.916 & 0.37 \\
$7oH$ & 110 & $N$&0.092//0.056 &0.976//0.992 &0.958//0.986 & 0.16//0.15 \\
$7gP$ & 180 &  ??&0.125//0.120&0.995//0.990 &0.978//0.987 & 0.282//0.145 \\
$10hP$ & 215.5 &$Y$&0.099 &0.943 &0.902 & 0.26\\
$10hH$ &110 &$N$ & ---//0.079 &---//0.992 &---//0.972 & ---//0.23\\
$10hZ$ &110 & $N$ &---//0.115&---//0.989 &---//0.974 & ---//0.34\\
$10oP$ & 514.3 &$Y$ &0.173 &0.977 &0.930 &0.39\\
$10oH$ & 110 &$N$ & 0.174//0.081 &0.964//0.981 &0.930//0.930 &0.12//0.30 \\  
$10gP$ & 294.6 &??&0.224//0.191&0.987//971&0.979//0.948& 0.172//0.158\\
\hline
\hline
\end{tabular}
\end{minipage}
\end{center}
\end{table*}

Our models show a wide variety of features, not only in the final state but also during the collision process. In Figures \ref{fig:j11}, \ref{fig:j12D} and \ref{fig:wrong}  we give some examples of the evolution seen in the various runs.

The first two, Figures \ref{fig:j11} and \ref{fig:j12D},  provide examples of the evolution of initial states leading to merger. Figure \ref{fig:j11} gives the evolution of a model with equal mass components in a rectilinear head-on collision (run $1hP$). A large fraction of the orbital energy is lost in the first encounter and rapid merging follows.

Next, Figure \ref{fig:j12D} shows the evolution of a system with mass ratio 2:1. The two galaxies are placed in a parabolic orbit with an impact parameter slightly smaller than the outer radius of the larger of the two galaxies (run $2gPr$). After the first encounter at $t=60$ the smallest galaxy features a tidal tail (`plume'); it also produces a small bridge of particles. The merger is complete around $t \approx 210$.

\begin{figure}

\begin{center}
\leavevmode
\hbox{%

\epsfxsize=8cm
\epsfbox{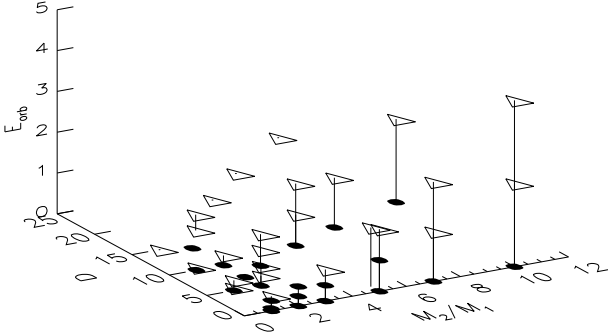}}
\caption{The space of input parameters: orbital energy, impact parameter ($D$) and mass ratio ($M_2/M_1$). All models of our sample are shown; those leading to merger are depicted as filled circles, non-mergers are depicted as open triangles. 
\label{fig:3d}}
\end{center}
\end{figure}

Finally, Figure \ref{fig:wr} shows an example of non-merging systems. The masses of the initial galaxies are 5:1 (run $5wW$). The systems are placed on a slightly non head-on hyperbolic orbit. This is not a typical case for non-merging systems in our sample, and we give it here as an illustration. The initial parameters are such that we are close to the border between mergers and non-mergers (see figure \ref{fig:3d}).   The most massive system is almost undamaged during the interaction. However, the small system loses particles that end up in the potential well of the bigger galaxy. Also a fairly narrow bridge between the two systems is formed. A small cloud of  particles leads the small galaxy as it runs away from its bigger companion (at the lower right-hand side of the frames).

The space of  the input parameters is given as a $3$-D plot in Figure \ref{fig:3d}.
Here the three coordinate axes are $E_{\rm orb}$, $D$ and $M_2/M_1$. Circles indicate mergers and triangles non-mergers. 
All the systems studied here involve intense galaxy interactions,  leading to several distinct features, also for those runs where merging does not occur. 

All runs (but one, see below) with $E_{\rm orb}=0$ in Table \ref{tab:Enohres} result in a merger. In these simulations the systems lose part of the orbital energy in the first encounters, which results in `heating' of the two galaxies. Subsequent passes through pericenter ultimately lead to merging. In all cases particles escape, carrying away some energy. We have considered a large impact parameter in only one case with $E_{\rm orb}=0$. For this run ($1wP$) the impact parameter is larger than the sum of the radii of the two systems;  ultimately it may result in a merger but the required time would be larger than a Hubble time. The same applies to all systems denoted with '??' in Table \ref{tab:Enohres}. Note that model $2gP$ takes a long time to merge (close to a Hubble time). This is why this model although considered as a merger is also marked with '??' in Table \ref{tab:Enohres}. The boundary between mergers and non-mergers in the plane $E_{\rm orb}=0$ in Figure \ref{fig:3d} has not been fully explored. 

\normalsize

Also some runs with energy $E_{\rm orb}>0$ lead to mergers. Those models have $D=0$, that is,  they are head-on collisions. For these the dynamical evolution is similar to that of the head-on collisions with $E_{\rm orb}=0$, but this is only true up to mass ratios  5:1.

In summary, mergers from the present sample without a dark halo are confined to a limited region in the 3-D space defined by orbital energy, impact parameter and mass ratio (see Figure  \ref{fig:3d}). For hyperbolic energies we have mergers only for head-on collisions and mass ratios close to one. 

\subsection{Morphology of the systems}

The various morphological features of the end products are summarized in Table \ref{tab:morf}. In the column `Class.' we give a `Hubble type' morphological classification for the merger remnants. To do so we have  fitted the isodensity contours by ellipses and we have taken the mean projected ellipticity inside $R_{\rm e}$ of the system from 100 random points of view (see Figure \ref{fig:typ}). Table \ref{tab:morf} gives the median value (note that in Table \ref{tab:Enohres} a value for the isophotal ellipticity at $R_{\rm e}$ for a given point of view is listed). The table also gives the kind of interaction. We considered the system to be a merger if more than half of the particles of each system are bound together in a single object as a result of the interaction. When this condition is not met we classify the interaction as follows. If more than $10 $ per cent of the particles of one system end up in the potential well of the other we call it `strong' interaction.  When there is an exchange of less than $10 $ per cent of the particles, it is denoted as `weak' interaction.

\begin{figure}
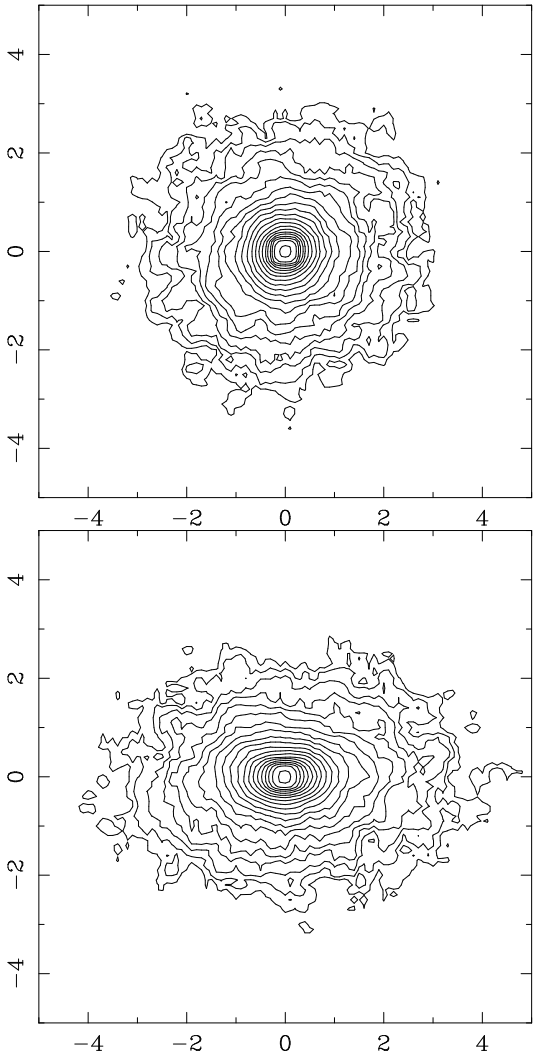


\centering
\includegraphics[width=7cm]{fig7_1.ps}
\hspace{0.cm}
\includegraphics[width=7cm]{fig7_2.ps}


\caption{Contour plots of the projected density distribution for the remnant in run $1hP$. This system, resulting from a head-on collision, has a prolate cigar-shape structure. Top view along the major axis, bottom view along short axis.}
\label{fig:cont1}
\end{figure}

\begin{figure}

\begin{center}
\leavevmode
\hbox{%

\epsfxsize=8cm
\epsfbox{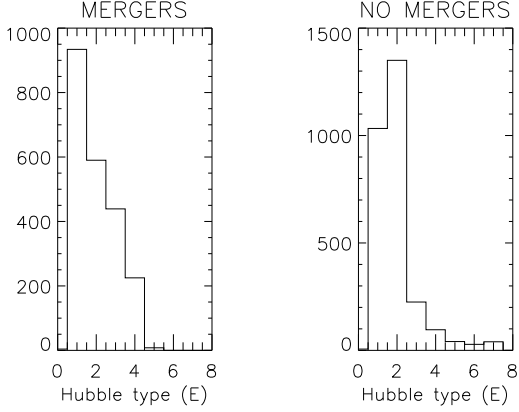}}
\caption{Distribution of mean ellipticity inside $R_{\rm e}$, calculated from one hundred random points of view. Left panel: results for merger end-products. Merger remnants cover the entire observed range of ellipticities, from E0 to E5. Right panel: results for non-merging systems. Note that the tails of both distributions are affected by `classification noise'.}
\label{fig:typ}
\end{center}
\end{figure}

The more flattened systems result from simulations with approximately equal mass components, roughly parabolic orbit and impact parameter smaller than $R_{\rm out}$ (see Figure \ref{fig:cont1} as an example). There is a clear trend from E5 to E1 as we go to higher mass ratios, as expected. Frequently the end product has type E3 (see Figure \ref{fig:typ}). In Figure \ref{fig:typ} we give the distribution of Hubble types for the final systems as calculated from the one hundred points of view. The left panel shows results for merger models, while the right panel shows non-mergers. In both cases we find a large range in ellipticities, from E0 to E5 for merger models and even higher for non-mergers. 

\begin{figure}

\centering
\includegraphics[width=5.5cm]{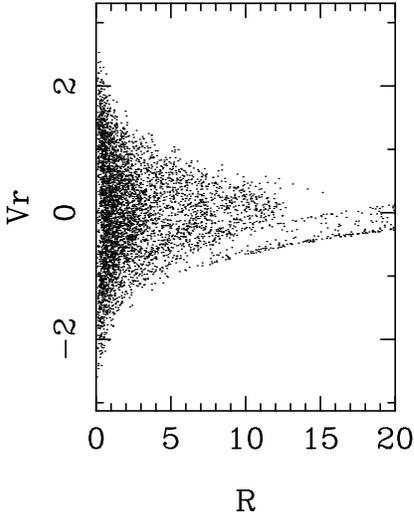}

\caption{In run $3hH$ with mass ratio 3:1 a number of faint shells are formed. This figure shows the radial velocity versus the radius for the particles belonging to the least massive galaxy, i.e. the progenitor of mass 1. The particles from this system are placed, via phase wrapping, in shells that can be seen in this figure as curved, relatively sharp features.}
\label{fig:shell3}
\end{figure}

For mergers resulting from systems with $E_{\rm orb} > 0 $ the least massive galaxy is usually highly distorted, some of the particles are trapped in the main potential well and shells are formed similar to those found by Hernquist \& Quinn (1988). Figure \ref{fig:shell3} gives the radial velocity versus the radius for the particles in the smaller galaxy for run $3hH$. Some of the particles from this system are placed on shells, visible as thin curved features.  Although these shells are less obvious than those found by Hernquist \& Quinn (1988) it should be noted here that these authors explored minor mergers between spheroids (mass ratios larger than 10:1) to produce shells. From figure \ref{fig:shell3} we find that shells are also produced in major mergers (3:1 in this case).

When looking at the non-merging systems we notice that the interaction is stronger when the masses are similar. However it seems that interaction features, such as bridges and tails, are more prominent if the masses are not equal, e.g. $M_2/M_1$ of order 3. 
For non-mergers with high orbital energy the most massive galaxy is almost undisturbed after the interaction, contrary to the least massive one. Part of its particles end up in the inner parts of its neighbour's potential well, while some are expelled and form a plume or `halo' around the escaping core (see Figure \ref{fig:wrong}). In models with $D$ close to zero a bridge may form.

\begin{table}
\begin{minipage}{85mm}
\caption{Morphological classification of the end products. The classification is indicative only; it refers to the value of the median when measuring the type from 100 random points of view. \label{tab:morf}}
\begin{tabular}{cccccc}
\hline
{\bf Run}&{\bf $M_2:M_1$}& {\bf $D$} & {\bf $E_{\rm orb}$} & {\bf Class.}& {\bf Interaction}\\
\hline
$1hP$ & 1:1 & 0 & 0 & E4 &merger \\
$1hH$ & 1:1 & 0 & 0.0625 & E4 &merger  \\
$1hZ$ & 1:1 & 0 & 0.250 &E4 &merger \\
$1hW$ & 1:1 & 0 & 0.360 & E1//E1 &strong\\
$1oP$ & 1:1 & 5 & 0 &E3 &merger\\
$1oH$ & 1:1 & 5 & 0.0625 &E2//E2  &strong \\
$1oZ$ & 1:1 & 5 & 0.250 & E2//E2 & strong\\
$1gP$ & 1:1 & 10 & 0 &E3 &merger \\
$1gPG$ & 1:1 & 10 & 0 &E3 &merger \\
$1gH$ & 1:1 & 10 & 0.0625&E2//E2 & strong\\
$1wP$ & 1:1 & 15 & 0 & E2//E2 &strong\\
$2hP$ & 2:1 & 0 & 0 & E3 &merger \\
$2hH$ & 2:1 & 0 & 0.239 &E3 &merger \\
$2hZ$ & 2:1 & 0 & 0.478 &E2 &merger \\
$2oP$ & 2:1 & 7.07 & 0 &E3 &merger \\
$2oPr$ & 2:1 & 5 & 0 &E3 &merger \\
$2oHr$ & 2:1 & 5 & 0.239 & E2//E1 & strong\\
$2oZr$ & 2:1& 5 & 0.478&E2//E1 & strong \\
$2gP$ & 2:1 & 14.14 & 0 &E2 &merger \\
$2gPr$ & 2:1 & 10 & 0 &E2 &merger \\
$2gHr$ & 2:1 & 10 & 0.239 & E2//E2 &strong \\
$3hP$ & 3:1 & 0 & 0 &E2  & merger \\
$3hH$ & 3:1 & 0 & 0.387 &E2 &merger \\
$3hZ$ & 3:1 & 0 & 0.775 &E4//E1 & strong\\
$3oP$ & 3:1 & 8.66 & 0 &E2 &merger\\
$3oH$ & 3:1 & 8.66 & 0.387 & E2//E2 &weak \\
$3oZ$ & 3:1 & 8.66 & 0.775 &E2//E1  &weak \\
$3gP$ & 3:1 & 17.32 & 0 &E2//E1 &strong \\
$3gH$ & 3:1 & 17.32 & 0.387 &E2//E1  &weak\\
$5hP$ & 5:1 & 0 & 0& E1&merger\\
$5hH$ & 5:1 & 0 & 0.761& E1&merger\\
$5hZ$ & 5:1 & 0 & 1.522&E3//E1 &strong \\
$5oP$ & 5:1 & 11.18 & 0&E1 &merger\\
$5oH$ & 5:1 & 11.18 & 0.761&E2//E1& strong\\
$5oZ$ & 5:1 & 11.18 & 1.522& E2//E1&weak\\
$5gP$ & 5:1 & 22.36 & 0 & E2//E1&weak\\
$5wW$ & 5:1 & 1.14 & 1.449&E0//E1&strong \\
$7hP$ & 7:1 & 0 & 0&E1&merger \\
$7hH$ & 7:1 & 0 & 1.22& E0//E0&weak\\
$7hZ$ & 7:1 & 0 & 2.44&E1//E0&weak\\
$7oP$ & 7:1 & 13.23 & 0&E1 &merger\\
$7oH$ & 7:1 & 13.23 & 1.22& E2//E1&weak\\
$7gp$ & 7:1 & 26.46 & 0&E2//E1& weak\\
$10hP$ & 10:1 & 0 & 0&E1&merger \\
$10hH$ & 10:1 & 0 & 2.039&E1//E0& weak\\
$10hZ$ & 10:1 & 0 & 4.078&E1//E0 &weak\\
$10oP$ & 10:1 & 15.81& 0&E1&merger\\
$10oH$ & 10:1 & 15.81 & 2.039&E2//E1& weak\\
$10gp$ & 10:1 & 31.62 & 0& E2//E1&weak\\
\hline
\end{tabular}
\end{minipage}
\end{table}


\subsection{Prolate and oblate systems}

We have measured the axial ratios $b/a$ and $c/a$ for each system, where  $a$, $b$ and $c$ are the principal axes of the ellipsoid. To do so, we have calculated the eigenvalues from the moment of inertia tensor:

\begin{equation}
I_{ij} = \int_V \rho({\bf r}) \left[ \delta_{ij} r^2 - x_ix_j \right] d^3{\bf x}.
\end{equation}

For a homogeneous ellipsoid with axes $2a$, $2b$ and $2c$, the eigenvalues of the inertia tensor are:

\begin{equation}
 E_1=(b^2+c^2)/5, \\\;\;
 E_2=(a^2+c^2)/5, \\\;\;
 E_3=(a^2+b^2)/5. \\\;\;
\end{equation} 

If we define the axes such that $E_1\leq E_2\leq E_3$, it follows that the axis ratios are:

\begin{equation}
	\frac{a}{b}=\sqrt{\frac{E_3+E_2-E_1}{E_1+E_3-E_2}}
\end{equation}

\begin{equation}
	\frac{a}{c}=\sqrt{\frac{E_3+E_2-E_1}{E_1+E_2-E_3}}
\end{equation}

\begin{figure}

\begin{center}
\leavevmode
\hbox{%

\epsfxsize=8cm
\epsfbox{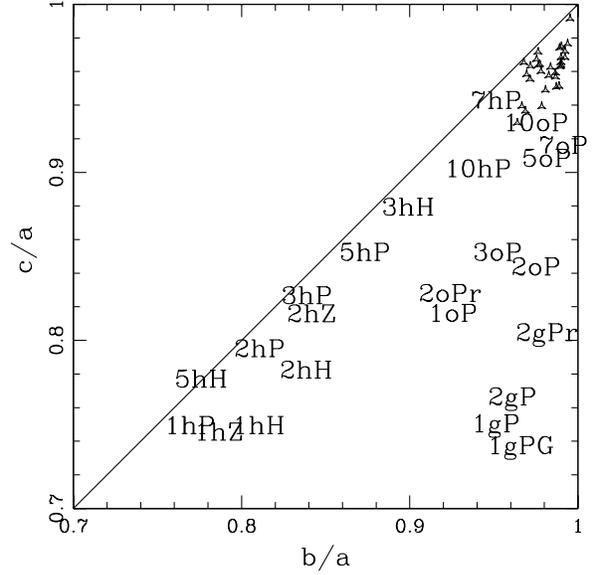}}
\caption{Axis-ratios of end-products in 'tri-axiality space' (de Zeeuw \& Franx 1991). Runs are identified as in Table \ref{tab1}. Triangles depict non-mergers.}
\label{fig:arat}
\end{center}
\end{figure}

Axis ratios for the merger remnants in our simulations have been calculated with these equations using particles whose distance to the center of mass is less than the half mass radius. Results are shown in Figure \ref{fig:arat}. The top right corner (1,1) represents a sphere, the lower right corner an oblate system and the lower left corner a prolate one (de Zeeuw and Franx 1991). Non-mergers are depicted as triangles and are nearly spherical as expected.

For our merger remnants, we find that head-on collisions ($D=0$) give nearly prolate end products, while for runs with $D\neq0$ the end products are triaxial or oblate spheroids. Merger remnants resulting from systems with mass ratios different from one are closer to spherical.

Our models achieve a flattening of up to $30 $ per cent of the initial system. It should be noted that the flattening achieved is not large, to some extend this may be caused by using the particles inside a sphere for the calculation of the axis ratios. Some elliptical galaxies are more flattened than the flattest systems produced in our runs.

\subsection{Rotation}

\begin{figure}

\begin{center}
\leavevmode
\hbox{
\epsfxsize=8cm
\epsfbox{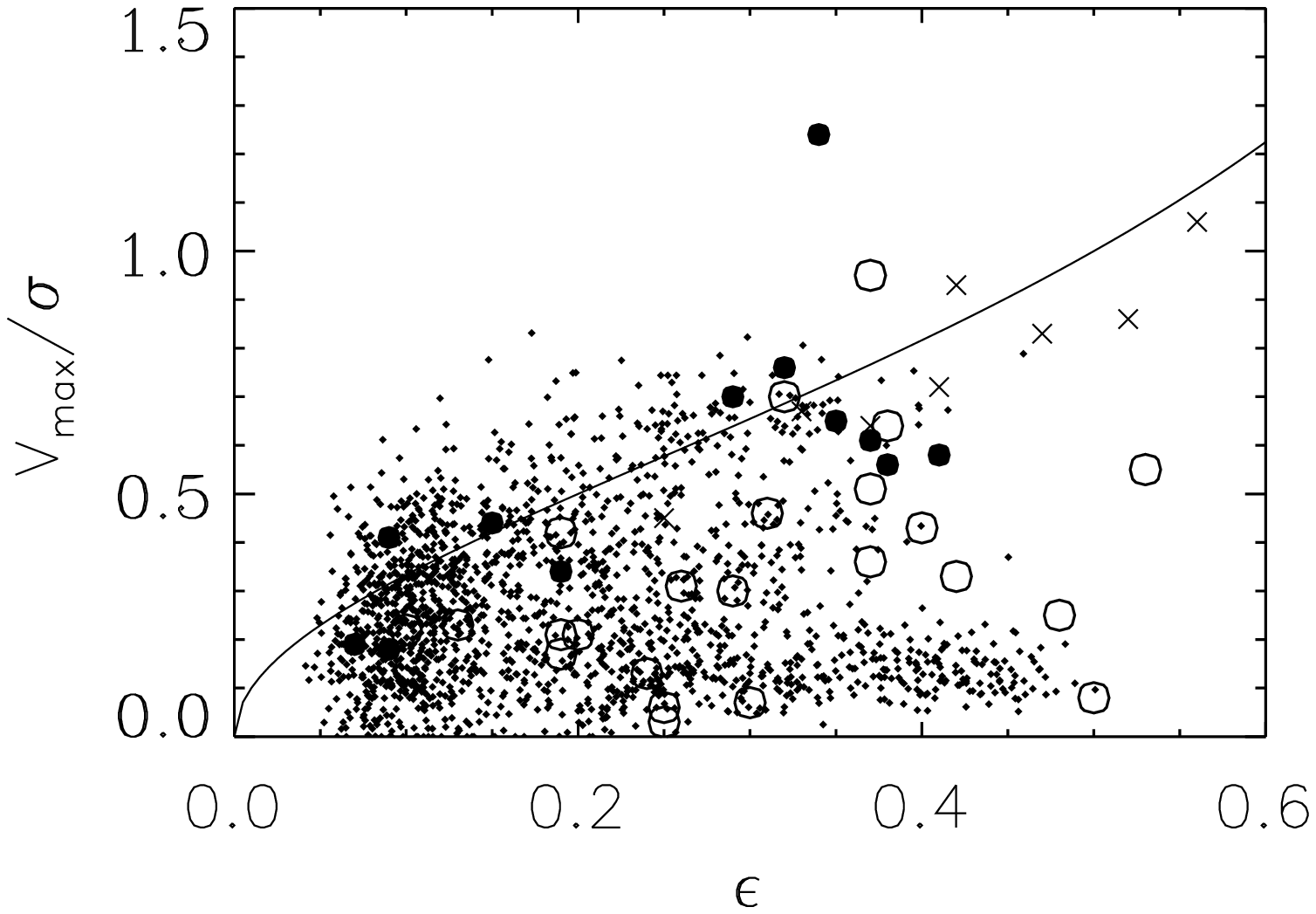}}
\hbox{%

\epsfxsize=8cm
\epsfbox{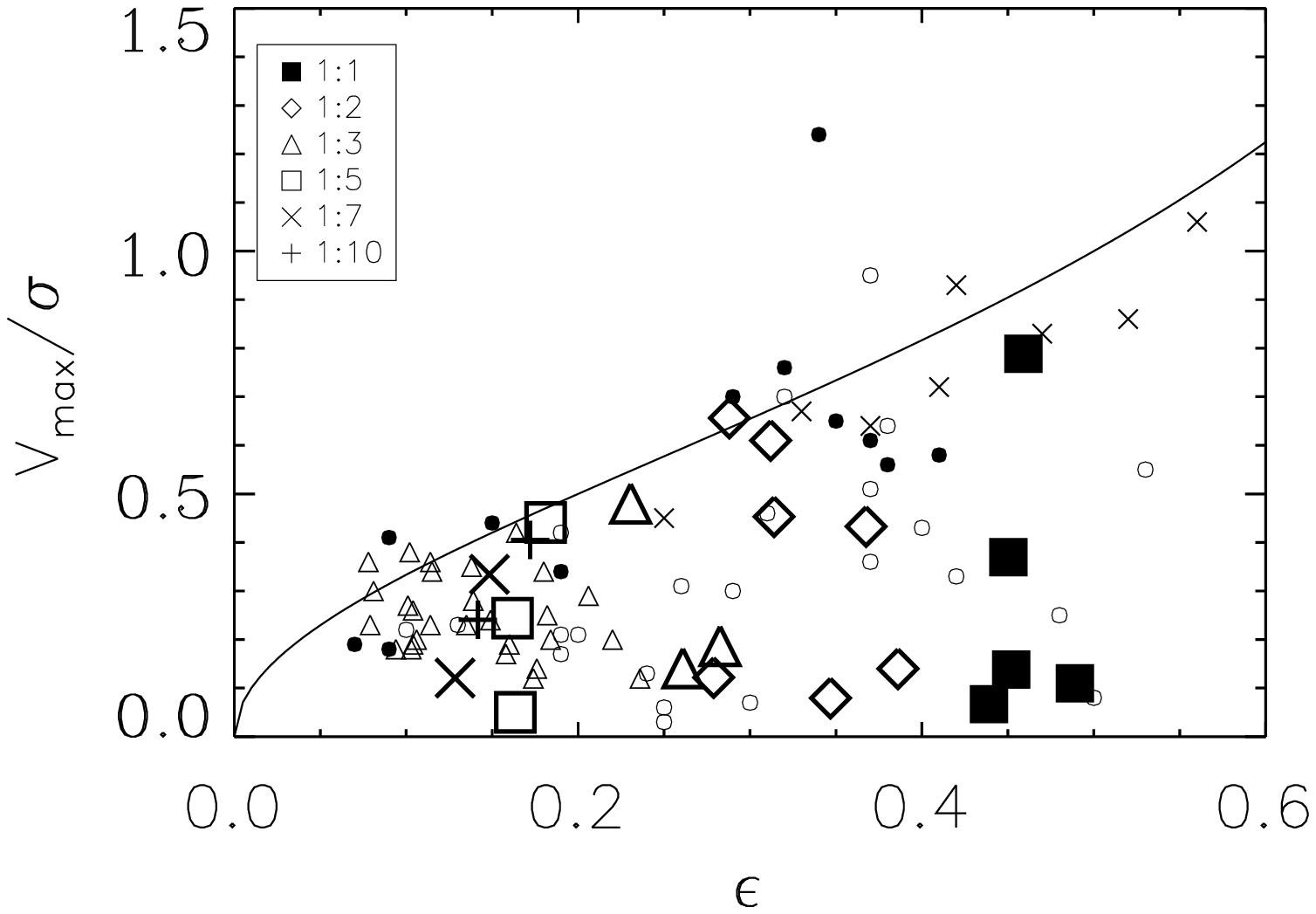}}
\caption{Streaming versus ellipticity for simulations compared with observational data. Open circles are high-luminosity ellipticals, filled circles are low-luminosity ellipticals while crosses are bulges (from Davies et al.~1983). Top panel shows the cloud of points (small dots) for the merger remnants and non-mergers measured from one hundred points of view.  Bottom panel shows the models as seen from a point of view along the y-axis of the initial configuration. From this point of view one recovers in general the maximal intrinsic flattening. Small symbols give the observational data, large symbols are the merger remnants and small triangles are the non-mergers. For details see text.}
\label{fig:evse}
\end{center}
\end{figure}

We have placed a slit along the major axis of the system as seen in projection for 100 randomly chosen points of view and we have measured the streaming velocity as a fuction of distance to the centre as well as the central velocity dispersion, $\sigma_{\rm o}$. Thus we measure the streaming velocity as a radial velocity along the line of sight, in a way that is similar to observations of ellipticals. $\sigma_{\rm o}$ was measured inside a radius of $0.2 \times R_{\rm e}$, i.e. about two softening radii (see also Gonz\'alez-Garc\'{\i}a and van Albada 2003). In Figure \ref{fig:evse} top we plot the ratio of the maximal streaming velocity, $V_{\rm max}$, and $\sigma_{\rm o}$ as a function of l.o.s. ellipticity at $R_{\rm e}$. Observational data are also given. Note that our merger remnants when seen from random directions are mainly small ellipticity systems with slow to mild rotation.

The last column in Table \ref{tab:Enohres} gives the ratio between the maximum streaming velocity and $\sigma_{\rm o}$ from a particular point of view perpendicular to the angular momentum vector. (This point of view corresponds in general to maximal flattening, that is, one sees the intrinsic flattening of the system.) This quantity has been plotted versus $\epsilonup$ measured from the same point of view, in Figure \ref{fig:evse} bottom. In this plot the merger remnants cover a large part of the space occupied by the observational data.  Non-mergers lie closer to the origin. 
Note that all equal mass mergers produce remnants with high intrinsic flattening. High values of $V_{\rm max}/\sigma_o$ is this figure are associated with large impact parameters. 

\subsection{Boxiness-disciness} 

For the end products of merger simulations we have calculated the deviation, ${\it \delta(\phi)}$, of the isodensity contours from pure ellipses and write:

\begin{equation}
	\delta(\phi) = \delta + \sum a_n {\rm \cos} (n\phi) + \sum b_n {\rm \sin} (n\phi).
\end{equation}

If the ellipse is a good description for the isophotes and it has been correctly fitted, $\delta$, $a_1,a_2,a_3$ and the $b_n$ should be small. Then, if the isophote is discy, $a_4$ will be positive and if it is boxy, $a_4$ will be negative (see Binney \& Merrifield 1998). To increase the signal to noise ratio  for our results an average over 60 configurations was made, always calculating the isophotes for a projection parallel to the intermediate-axis. These 60 configurations are different `snapshots' obtained by evolving the final systems a bit further.

\begin{figure}

\begin{center}
\leavevmode
\hbox{%

\epsfxsize=8cm
\epsfbox{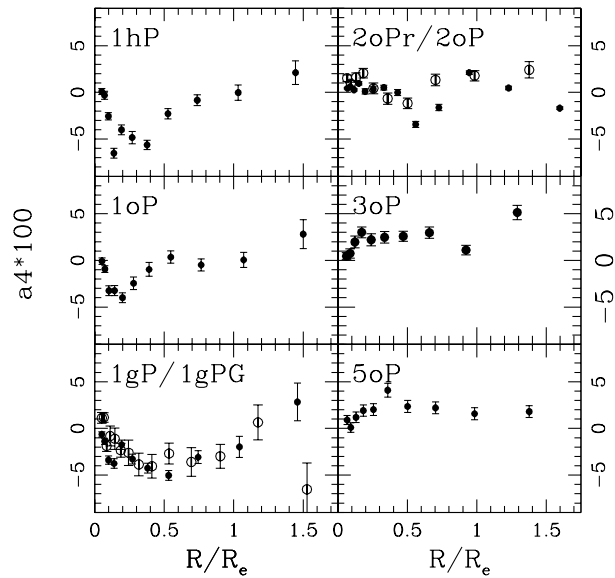}}
\caption{Mean radial variation of $a_4$ after equilibrium has been reached (based on $60$ snapshots along the intermediate axis) for runs identified in the upper left corner of each panel. Panels on the left show a boxy structure inside $R_{\rm e}$, those on the right a discy structure. Where two models are shown on the same frame the first one is given by filled circles and the second by open ones. For details see text.}
\label{fig:boxy2}
\end{center}
\end{figure}

Figure  \ref{fig:boxy2} shows the variation of $a_4$ with radius for seven merger models. Left panels show models with mass ratios $1:1$, parabolic orbit, and different impact parameters. For all those systems the ellipsoid is boxy inside $R_{\rm e}$. 
These three systems have a variety of characteristics. The top panel refers to a non-rotating radially anisotropic system, while the other two are anisotropic rotators (see section \ref{sec:isot}).  The bottom left panel also shows the results for the run $1gPG$ with an average for 6 snapshots, the results are quite similar to run $1gP$.

 The right panels in Figure \ref{fig:boxy2} show the $a_4$ parameter for four models with non-equal masses. They have in common that the impact parameter of their orbits is equal to half the radius of the larger initial system. In these cases the end products have a mildly discy structure. This disciness is more evident for the models with mass ratio 3:1 (run $3oP$, middle panel) and 5:1 (run $5oP$, bottom panel).

We also find, in agreement with Stiavelli et al.~(1991) and Governato et al.~(1993) that boxiness-disciness is subject to changes in the appearance of the same system due to projection effects: the same system can be seen as discy or boxy depending on the point of view. Our 1:1 models are mainly developing boxy isophotes. 

It is important to notice that this process of collisionless non-equal mass merging with an impact parameter $D \neq 0$ gives on average discy isophotes (see Figure \ref{fig:boxy2}, right panels). Although perhaps unexpected, it can be easily explained in terms of transfer of orbital angular momentum. The small system brings part of the orbital angular momentum to the inner regions of the remnants, transferring this into internal spin angular momentum. The smaller galaxy thus disrupts and preferentially populates a plane perpendicular to the angular momentum vector, giving rise to discy isophotes. In an ideal case one could think of this process as a way to form highly flattened spheroids inside dark matter haloes simply by the encounter of a spherical luminous system with a dark matter halo  which in that case must dominate the mass distribution. (The merged system would then have a high $M/L$ value). We find the same process acting in model $5gP$, described in section 3.7.

\subsection{Figure rotation and velocity anisotropy}
\label{sec:isot}

\begin{figure}

\begin{center}
\leavevmode
\hbox{%

\epsfxsize=7cm
\epsfbox{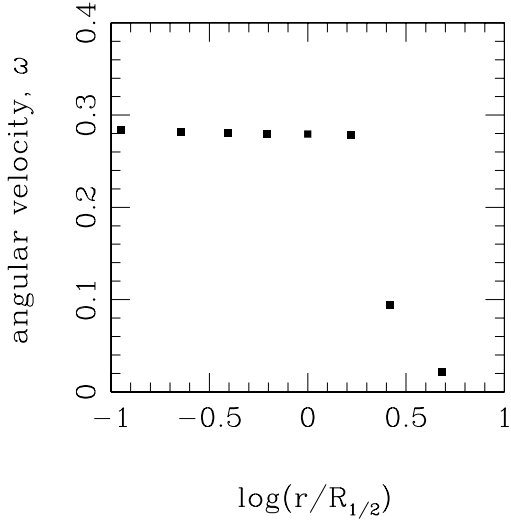}}
\caption{Pattern velocity as a function of radius for the end product of run $1oP$. A bar is present that extends beyond the half-mass radius, and has a sharp cut-off at twice the half-mass radius.}
\label{fig:tumb}
\end{center}
\end{figure}

With the algorithm outlined in section 3.4, we also calculated the eigenvectors of the inertia tensor. We used these to study the evolution of figure rotation for the merger models up to 20 dynamical times after merging. This has been done for the rotating systems only. As illustrated in Figure \ref{fig:tumb}, the results show that, in addition to particle streaming there is a bar-like rotating figure extending to about two half-mass radii. The bar is more prominent for low mass ratio encounters (1:1, 2:1).

\begin{figure}

\begin{center}
\leavevmode
\hbox{%

\epsfxsize=8cm
\epsfbox{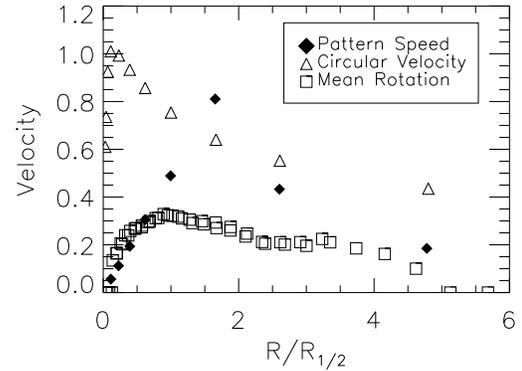}}
\caption{Streaming and figure rotation in the end product of run $1oP$. Triangles: circular velocity from the potential, open squares: streaming velocity, and diamonds: pattern speed.}
\label{fig:vrt}
\end{center}
\end{figure}

To confirm this point we have plotted the various relevant properties versus radius in Figure \ref{fig:vrt}. The upper curve represents the circular velocity calculated from the potential. The lower two represent the streaming velocity from two points of view perpendicular to the angular momentum vector. The pattern speed is indicated by the diamonds. The corresponding angular velocity is nearly constant inside $R/R_{1/2}=2$. 

The initial systems have isotropic velocity distributions. The end products of the merger process are often anisotropic however.
A measure of this quantity is given by the anisotropy parameter, $\beta$, defined as:

\begin{equation}
	\beta = 1-\frac{\sigma_{\rm t}^2}{2 \sigma_{\rm r}^2},
\end{equation}
where $\sigma_{\rm t}$ is the tangential velocity dispersion:

\begin{equation}
	\sigma_{\rm t}^2=\sigma_{\rm \theta}^2 + \sigma_{\rm \phi}^2.
\end{equation}

\begin{figure}

\begin{center}
\leavevmode
\hbox{%

\epsfxsize=9cm
\epsfbox{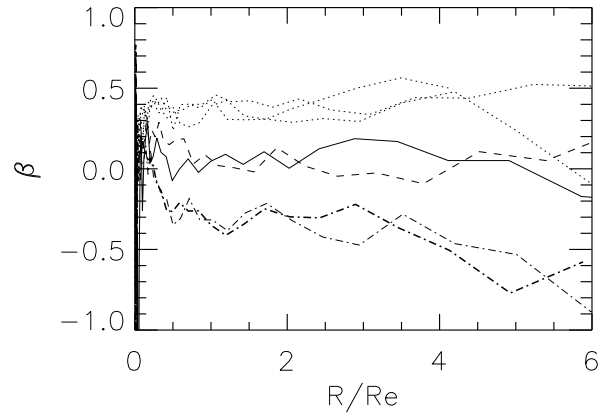}}
\caption{The anisotropy parameter $\beta$ for the mergers with equal mass progenitors. The solid line gives the initial model, dotted lines are models with $D=0$, dashed lines and  dashed-dotted lines represent models with $D \neq 0$ ($1oP$ and $1gP$ respectively). The thick dashed-dotted line represents model $1gPG$.\label{fig:aniso}}
\end{center}
\end{figure}

We have measured this anisotropy parameter for models with mass ratio 1:1 leading to a merger. The results are shown in fig. \ref{fig:aniso}.

Systems resulting from a head-on collision develop a radial anisotropy ($\beta>0$). Models with non-zero impact parameter show two types of behaviour. For small impact parameters, therefore small orbital angular momentum, we find no net anisotropy. For large impact parameter, i.e. large orbital angular momentum, the systems develop a tangential anisotropy, ($\beta < 0$). 

Thus, the shape of the orbit, that is, the initial orbital angular momentum, is reflected in the final orbital structure of the stars. 

\subsection{Noteworthy features}

\begin{figure}
\begin{center}
\leavevmode
\hbox{%
\epsfxsize=7cm
\epsfbox{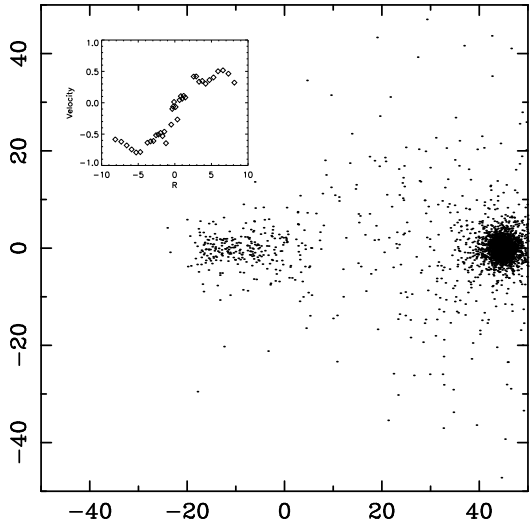}}
\caption{In due time, simulation $5gP$  probably will end up in a merger. The figure shows an intermediate state when an interesting situation occurs; only particles from the least massive system are shown. Part of the particles have left their parent system. They now lie in the potential well of the more massive galaxy (not shown here, centered at (-10,0)) and form a rotating disc as seen in the top left inset.}
\label{fig:cont5}
\end{center}
\end{figure}

Two interesting cases are runs $5gP$ and $10gP$. The systems in these runs will probably merge in the end, but the evolution time is so long that the simulation was stopped long before. At the first peri-centre passage a small fraction (around 2 per cent) of the least massive galaxy was stripped off and these particles settled into a disc inside the more massive galaxy, forming a rotating disc (see Figure \ref{fig:cont5}). 

\section{Discussion}

\parskip=0.6em
We have presented results of merger simulations of spherical, one-component systems, varying three parameters: orbital energy, impact parameter and mass-ratio. Dark matter has been omitted at this stage. In a following paper (Paper II) we address the differences between systems with and without dark matter haloes.

Whether or not a merger takes place is determined by the initial conditions and our `merger space' is in principle similar to that given by Binney \& Tremaine (1987, Figure 7-9). The structure of the merger remnants, which show a wide variety of morphological and kinematical characteristics, also reflects the values of the initial parameters. We find prolate non-rotating ellipsoids, as well as oblate systems with considerable rotational support. Our principal conclusions are as follows:

\begin{itemize}

\item The properties of the merger remnants match many of the global features of elliptical galaxies.

\item The initial orbital angular momentum defines the shape and the orbital structure of the remnant to a high degree. Head-on encounters produce prolate spheroids; off-axis encounters tri-axial or oblate systems.

\item Tumbling systems or oblate systems containing a rotating bar may form in
collisions with high impact parameter and mass-ratio close to 1.

\item Remnants resulting from nearly equal mass mergers can be appreciably flattened and are either non-rotating (consistent with luminous ellipticals) or flattened oblate rotators.

\item Off-axis encounters of ellipticals with mass ratio different from 1 can result in the formation of a rotating disc of stars in the inner parts of the main merger remnant (due to material stripped away from the smaller galaxy during the encounter).

\item Boxy as well as discy deviations in the isophotes are obtained. Discy deviations are found especially in models with impact parameter $ D \neq 0$. Disciness-boxiness is point of view dependent.

\item Shells may form in collisions involving systems with a large difference in mass, in accordance with Quinn (1984), Hernquist \& Quinn (1988) and Hernquist \& Spergel (1992). We also observe the formation of shells for mergers of mass ratio 3:1.

\end{itemize}

Obviously, further study is needed of the probability of finding initial
configurations as studied here in `real-life' systems, for instance by studying configurations in cosmological simulations. This would give a measure of the
likelihood of `gas-free' merging for the formation of ellipticals as seen today. 

In fact, observational evidence pointing towards merging of spheroidal systems as an important formation mechanism for ellipticals is mounting. Bell et al.~(2003) study over 5000 early-type galaxies at redshifts up to 1. They find that in their sample there are few bright blue galaxies that may fade into the luminous red galaxies (early types) seen today, and argue that luminous red galaxies (E's) could be
the outcome of mergers between less luminous bulge-dominated galaxies. Van Dokkum \& Ellis (2003) observe traces of star-formation for a sample of
galaxies in the Northern Hubble Deep Field. The mass associated with these regions of star formation seems to be small however and they suggest
the possibility of mergers between bulge-dominated, gas-poor systems as a
likely explanation. Further, Khochfar \& Burkert (2003), from semi-analytical modelling, find
that the fraction of mergers of bulge dominated galaxies (early types)
increases with time, independently of environment. They also conclude that
at least $50\%$ of present day E's are relics of major mergers between
spheroidal systems.

Elliptical galaxies may be divided into two broad groups concerning their
kinematics and orbital structure. One group contains the low-luminosity
elliptical galaxies that are rotationally supported and have discy
deviations of their isophotes. The second group contains the high-luminosity
elliptical galaxies that are radially anisotropic,
slow rotators with boxy isophotes. Clearly, because the merger remnant models can be scaled to any size, our merger simulations can not explain such relations; they lack an essential element.

The issue boxyness vs. disciness is complicated and we will not speculate on the possible interpretation of the observed correlations (Nieto et al.~1994, Bender et al.~1994, Kormendy \& Bender~1996, Mihos \& Hernquist 1996, Faber et al.~1997, Naab et al.~1999). Here we want to emphasize that mergers can produce both boxy and discy isophotes. The crucial parameter is the impact parameter: off-axis encounters produce mainly (oblate) merger remnants with discy isophotes while head-on collisions produce prolate merger remnants with boxy isophotes. But these results are complicated by the point of view dependence of boxyness and disciness.

For the observables explored here, the formation of elliptical galaxies may need no disc intervention at all. This result would extend the conventional picture of hierarchical merging where from assembly of complicated (disc) systems we would get to simple (E0-E1) systems, as well as
more highly flattened systems. The present study shows that from nearly
spherical systems one may also get more complicated flattened spheroidal 
systems.

Mergers between spheroidal systems are likely to happen and as shown here they do not erase the characteristics of E's. For example, the intrinsic shape of the remnant can be directly related to the impact parameter of the collision. In other words, prolateness and oblateness are probably relics of the orbital parameters. Mergers keep a memory of the dynamics of the encounter and some characteriscis of ellipticals can be built during the merging stage. Mergers involving disks and gas introduce a broader range of features, but E+E mergers may provide a way of producing many (although not all) observed characteristics of E's that should not be overlooked.

\end{document}